\documentclass[aps,prl,twocolumn,superscriptaddress,preprintnumbers,floatfix,tightenlines,reprint
]{revtex4-1}

\pdfoutput=1
\usepackage{graphics}
\usepackage{natbib}
\usepackage{amsfonts,amsmath,amssymb}
\usepackage{graphicx}
\usepackage{epstopdf}
\usepackage{slashed}
\usepackage{bm}
\usepackage{tikz}
\usetikzlibrary{matrix}
\usetikzlibrary{shapes.arrows,chains}
\usepackage{hyperref}

\def\CA{{\cal A}}

\def\CK{{\cal K}}

\def\CN{{\cal N}}

\def\a{\alpha}

\def\dd{d}

\newcommand{\frakM}{\mathfrak{M}}

\newcommand{\be}{\begin{equation}}
\newcommand{\ee}{\end{equation}}

\newcommand{\hatM}{\hat{M}}
\newcommand{\bhbar}{\hbar}

\def\Tr{{\rm Tr}}

\hypersetup{
    colorlinks=true,       
    linkcolor=red,          
    citecolor=blue,        
    filecolor=magenta,      
    urlcolor=blue           
}

\begin{document}

\preprint{IPMU15-0171}

\title{Taming Supersymmetric Defects in 3d--3d Correspondence}
\author{Dongmin Gang}

\affiliation{Kavli Institute for the Physics and Mathematics of the Universe (WPI),
University of Tokyo, Chiba 277-8583, Japan}
\author{Nakwoo Kim}

\affiliation{Department of Physics and Research Institute of Basic Science,
Kyung Hee University, Seoul 02447, Korea}
\affiliation{School of Physics, Korea Institute for Advanced Study, Seoul 02455, Korea}
\author{Mauricio Romo}

\affiliation{Kavli Institute for the Physics and Mathematics of the Universe (WPI),
University of Tokyo, Chiba 277-8583, Japan}

\affiliation{School of Natural Sciences, Institute for Advanced Study, Princeton, NJ 08540, USA}
\author{and Masahito Yamazaki}

\affiliation{Kavli Institute for the Physics and Mathematics of the Universe (WPI),
University of Tokyo, Chiba 277-8583, Japan}

\affiliation{School of Natural Sciences, Institute for Advanced Study, Princeton, NJ 08540, USA}

\begin{abstract}
We study knots in 3d Chern-Simons theory with complex gauge group $SL(N,\mathbb{C})$,
in the context of its relation with 3d $\mathcal{N}=2$ theory (the so-called 3d--3d correspondence).
The defect has either co-dimension 2 or co-dimension 4 inside the 6d $(2,0)$ theory,
which is compactified on a 3-manifold $\hat{M}$.
We identify such defects in various corners of the 3d--3d correspondence, namely
in 3d $SL(N,\mathbb{C})$ Chern-Simons theory, in 3d $\mathcal{N}=2$ theory, in 5d $\mathcal{N}=2$ super Yang-Mills theory, and in the M-theory holographic dual. We can make quantitative checks of the 3d--3d correspondence
by computing partition functions at each of these theories.
This Letter is a companion to a longer paper \cite{GKMY_long}, which contains more details and more results.
\end{abstract}

\maketitle

{\bf Introduction.}---One lesson from history is that physics and mathematics often develop hand in hand; the development on one side
facilitates development in the other, creating a virtuous cycle of feedback.  The recently-discovered
3d--3d correspondence \cite{Terashima:2011qi,Terashima:2011xe,Dimofte:2010tz,Dimofte:2011jd,Dimofte:2011ju,Cecotti:2011iy}
is a perfect example for this interplay. The correspondence
states that there exists a surprising connection between 3d $SL(N,\mathbb{C})$ Chern-Simons (CS) theory defined on a 3-manifold $M$
on the one hand, and a 3d $\CN=2$ supersymmetric gauge theory (which we call $T_N[M]$) on the other.
Being a topological field theory, the $SL(N,\mathbb{C})$ CS theory provides a functor equipped with a Hilbert space. For the $N=2$ case, we have the
$SL(2,\mathbb{C})$ CS theory and the relevant Hilbert space corresponds to a quantization of the
moduli space of $SL(2,\mathbb{C})$-flat connections on $M$, and it contains the space of hyperbolic structures on $M$ when $M$ is a hyperbolic manifold.
The 3d-3d relation thus unifies and enriches the mathematics of both knot theory and 3d hyperbolic geometry.

The physical origin of the 3d--3d correspondence is the
compactification of the 6d $(2,0)$ theory on a closed 3-manifold $\hatM$,
along which the theory is partially topologically twisted:
\begin{align}
\textrm{ $N$ M5s on } \mathbb{R}^{1,2} \times \hatM \label{(2,0) on hatM}\;.
\end{align}
The 3d $\mathcal{N}=2$ theory $T_N[\hatM]$ lives on $\mathbb{R}^{1,2}$, which is transverse to $\hatM$.

\begin{figure}[htbp]
\begin{center}
   \includegraphics[width=.18\textwidth]{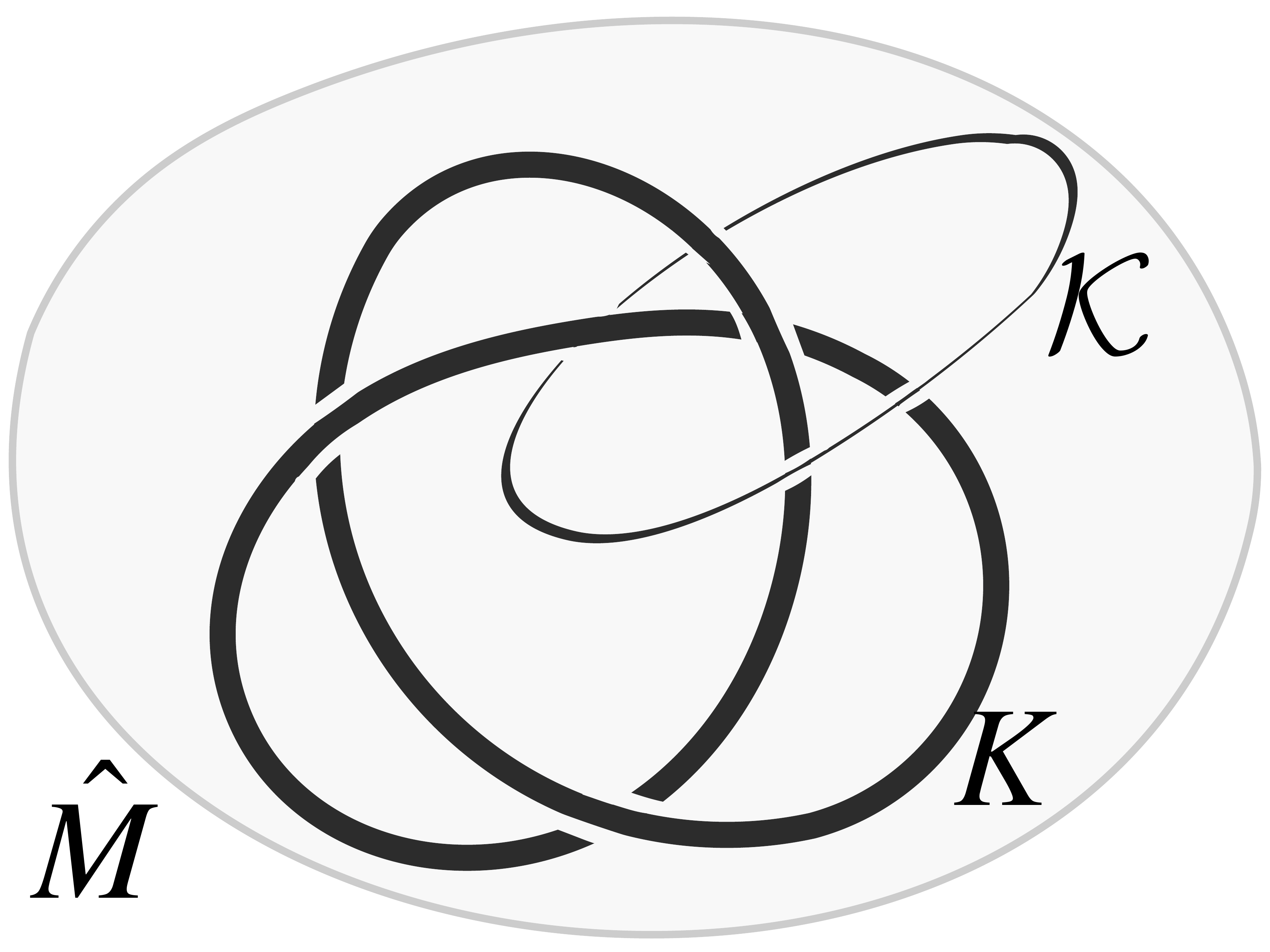}
   \end{center}
   \caption{We have knot-like defects inside a closed 3-manifold $\hatM$. We in general simultaneously include a co-dimension $2$ defect along $K$, and then
   a co-dimension 4 defect along $\CK$. The two knots, $K$ and $\CK$, can be knotted inside $\hatM$.}
    \label{fig:K_CK}
\end{figure}

{\bf Loop-like Defects.}---In this Letter we study the inclusion of supersymmetry-preserving defects to the 3d--3d correspondence.
The defects fill in knots (or links) inside the closed 3-manifold $\hatM$, see Fig.~\ref{fig:K_CK}.

There are two types of defects. They are easier to understand in terms of the M5-brane configuration,
where the defect has either 2 or 4 co-dimensions:
\begin{align}
\begin{array}{cccc|ccc|ccccc}
&  \multicolumn{3}{c}{\overbrace{ }^{\mathbb{R}^{1,2}}}    &  \multicolumn{3}{c}{\overbrace{ }^{\hatM}}   &  &  &
\\
\textrm{$N$ M5 } & 0 & 1 & 2 & 3 & 4 & 5 &   &  &  &  &
\\
\textrm{(Co-dim.\ 2) M5 } & 0 & 1 & 2 & 3 & &  &   & 7 & 8
\\
\textrm{(Co-dim.\ 4) M2 } & 0 &  &  & 3 & &  & 6  &  &  & &
\\
\textrm{(Co-dim.\ 4) M5 } & 0 &  &  & 3 & &  &   & 7 & 8 & 9 &\sharp
\end{array}
\label{codim_4_brane}
\end{align}
The co-dimension 4 defect is described either by the M2-brane or its blow-up,  the M5-brane (more comment on this later).
We will include a co-dimension 2 defect along a knot $K$, and and co-dimension 4 along another knot $\CK$.

There are several motivations for studying such defects in the 3d--3d correspondence.
First, co-dimension 2 defects can be thought of as cutting out a tubular neighborhood of $K$ inside $\hat{M}$ and prescribe a particular holonomy along a cycle
of the boundary torus. These are well-studied in the mathematical literature on knot theory and hyperbolic geometry ({\it e.g.}\,\,\cite{ThurstonLecture}),
and in fact most of the discussions on the 3d--3d correspondence in the literature
already contain such a co-dimension 2 defect $K$. The 3-manifold $M$ is then taken to be
a complement (exterior) of a knot inside a closed 3-manifold $\hatM$:
\begin{align}
M = \hat{M} \backslash K  \;,
\label{M_def}
\end{align}

Second, despite their importance, not much is known about these defects.
For example,  when we consider the case of $N>2$ M5-branes, a co-dimension $2$ defect along $K$
could be  of a ``non-maximal'' type (as we will discuss later), and almost nothing is known about
these cases. The resulting partition function, with the defects included, will be a quantity of both mathematical and physical interest,
and gives a new invariant of a knot, in particular.

Finally, the defects help us to better understand the proposed 3d $\CN=2$ theory
$T_N[M]$. In the literature, we find that there have been two different types of proposals for the the
$T_N[M]$ theory, using either Abelian or non-Abelian gauge groups.
The considerations of our Letter and the forthcoming paper \cite{GKMY_long} makes it clear that we need a non-Abelian description for the
proper understanding of co-dimension 4 defects.

In the rest of this Letter, let us discuss our defects in
various different theories appearing in the 3d--3d correspondence:
\begin{align}
\textrm{
\scalebox{0.9}{\begin{tikzpicture}[
node distance=2.4cm,
roundnode/.style={circle, draw=green!60, fill=green!5, very thick, minimum size=5mm},
squarednode/.style={rectangle, draw=red!60, fill=red!5, very thick, minimum size=5mm},
TSUNnode/.style={circle, draw=blue!60, fill=blue!5, very thick, minimum size=5mm},
]
\node[squarednode](N){6d $(2,0)$ theory};
\node[squarednode](NE)[above right of=N]{holographic dual};
\node[squarednode](W)[below left of=N]{3d $\CN=2$ theory $T_{N}[M]$};
\node[squarednode](E)[below right of=N]{3d $SL(N)$ CS};
\node[squarednode](NW)[above left of=N] {5d $\CN=2$ SYM};
\draw(N) -- (W);
\draw(N) -- (NE);
\draw(N) -- (NW);
\draw(N) -- (E);
\end{tikzpicture}
}
}
\label{corners}
\end{align}

{\bf 5d.}---Let us begin with 5d $\CN=2$ $SU(N)$ super-Yang-Mills (SYM). The advantage of this theory, when compared with the 6d $(2,0)$ theory,
is that the theory has an explicit Lagrangian.
In fact, when we replace $\mathbb{R}^{1,2}$ in \eqref{(2,0) on hatM} by an $S^1$-bundle over $S^2$, we can reduce the 6d theory along the $S^1$,
to obtain 5d $\CN=2$ SYM on $S^2\times \hat{M}$.
We can then perform the supersymmetric localization computation
and show directly that the theory on $\hat{M}$ is the 3d $SL(N,\mathbb{C})$ CS theory \cite{Cordova:2013cea,Lee:2013ida} (see also \cite{Yagi:2013fda}).

We can discuss the supersymmetric defects in this setup.
A co-dimension 2 defect is realized by coupling the 5d $\CN=2$ SYM with the
so-called $T_{\rho}[SU(N)]$ theory \cite{Gaiotto:2008sa,Gaiotto:2008ak} ({\it cf.}\,\,\cite{Bullimore:2014upa,Yonekura:2013mya}).
Here $\rho$ is an embedding $\rho: \mathfrak{su}(2) \rightarrow  \mathfrak{su}(N)$, or equivalently a partition of $N$:
\begin{align}
\rho = [n_1, n_2, \ldots, n_s]\;,\quad n_i\ge n_{i+1}\;, \quad \sum_{i=1}^s n_i=N \;.
\label{rho}
\end{align}
The defect is called `maximal' (`simple') when
$\rho=[1]^{N}\!:=[1,1,\ldots, 1]$ ($\rho=  [N-1 ,1]$).
The $T_{\rho}[SU(N)]$ theory has flavor symmetry $SU(N)\times H_{\rho}$, where $H_{\rho}$ is the commutant of the image of $\rho(SU(2))$ inside $SU(N)$.
The 5d theory couples to $T_{\rho}[SU(N)]$ by gauging the $SU(N)$ part of this flavor symmetry.

A co-dimension 4 defect is realized by a supersymmetric Wilson loop in 5d $\CN=2$ SYM:
\begin{align}
\begin{split}
Z_R^{\rm 5d}
&=\langle W_R \rangle\\
&= \left\langle \textrm{Tr}_R \, P \exp\left( \int_{ \{p \}\times \CK} \left(-A_{\mu} \mp i \phi_{\mu}\right) d\tau^{\mu} \right)  \right\rangle \;.
\end{split}
\label{5d_W}
\end{align}
Here $\phi_{\mu}$ are three of the adjoint scalar fields of the 5d $\CN=2$ SYM, which after the topological twist are turned into
a 1-form on $M$. The point $p\in S^2$ should be either the north or the south pole in order to preserve some supersymmetry.
The co-dimension 4 defects are hence labeled by
\begin{align}
R\; :\; \textrm{a unitary representation of $SU(N)$}\;.
\label{R}
\end{align}

{\bf 3d Chern-Simons.}---Let us next consider the theory on $M$, the 3d $SL(N,\mathbb{C})$ CS theory \cite{Witten:1989ip}.
The Lagrangian of the complexified Chern-Simons theory  is given by
\begin{align}
S_\textrm{CS}[\CA,\overline{\CA};\mathbf{\bhbar},\tilde{\bhbar}]
=\frac{i}{2\bhbar} \textrm{CS}[\CA]+\frac{i}{2\tilde{\bhbar}} \textrm{CS}[\overline{\CA}]\;\;, \label{complex CS action}
\end{align}
where the CS functional defined by
\begin{align}
\textrm{CS}[\mathcal{A}]:=\Tr \left( \mathcal{A}\wedge d\CA + \frac{2}3 \CA \wedge \CA \wedge \CA \right)\;,
\end{align}
and we defined in general complex ``Planck constants'' by
\begin{align}
\bhbar: =\frac{4\pi i}{k+\sigma} \;, \quad \tilde{\bhbar}: =\frac{4\pi i }{k-\sigma} \;,
\label{Planck}
\end{align}
with $k\in \mathbb{Z}$ and $\sigma \in  \mathbb{R}  \textrm{ or } i\mathbb{R}$.
Mathematically, this theory gives a quantization of the moduli space of $SL(N,\mathbb{C})$-flat connections on $M$.

In CS theory, a co-dimension 2 defect corresponds to a monodromy defect, which is to specify the holonomy along the boundary torus of the complement of $K$.
More precisely, the manifold \eqref{M_def} has a boundary torus
\begin{align}
\partial \left(\hat{M}\backslash K \right) = T^2 \ .
\end{align}
The torus has two non-contractible cycles, the meridian $\mathfrak{m}$ (contractible in the tubular neighborhood of $K$)
and the longitude $\mathfrak{l}$.
In quantum theory it suffices to specify the holonomy for one of them, and the boundary holonomy along the meridian $\mathfrak{m}$ is taken
to be
%
\begin{align}
\textrm{Hol}_{\mathfrak{m}}(\mathcal{A}) \sim  \left(\begin{array}{cccc}e^{\frakM_1}\mathbb{I}_{n_1} & 0 & \bf{0} & 0 \\0 & e^{\frakM_2}\mathbb{I}_{n_2} & \bf{0} & 0 \\ \bf{0} & \bf{0} & \ldots & \bf{0} \\0 & 0 & \bf{0} & e^{\frakM_s}\mathbb{I}_{n_s} \end{array}\right) \,,
 \label{co-dimension 2 defects as b.c. in CS theory-1}
\end{align}
where the size of the block is determined by the partition $\rho$ \eqref{rho},
and $\sim$ denotes equivalence under the adjoint action of the gauge group.
The partition function is defined by the path-integral
\begin{align}
Z_{\rho}^\textrm{CS}= \int \, [\mathcal{D}\mathcal{A}] [\mathcal{D} \overline{\CA}] \,\, e^{i  S_\textrm{CS}[\CA,\overline{\CA};\bhbar,\tilde{\bhbar}] } \,,
\label{Z_CS}
\end{align}
with the boundary condition along $K$ as in \eqref{co-dimension 2 defects as b.c. in CS theory-1}.
The path-integral \eqref{Z_CS} can be defined for arbitrary coupling constants $\hbar,\tilde{\hbar}$ by enlarging the domain of integration 
 and deforming the integration contour \cite{Witten:2010cx}, as we will comment more in \cite{GKMY_long}.
The existing literature has focused on the case of maximal puncture, see in particular \cite{BFG,2011arXiv1111.2828G,GGZ,Garoufalidis:2013upa,Dimofte:2013iv} for the case $N>2$.

On the other hand, a co-dimension 4 defect in 3d CS theory is a Wilson line along the knot $\mathcal{K}$ in representation $R$ (more precisely, in $SL(N,\mathbb{C})$ CS theory $R$ becomes the natural holomorphic lift of the $SU(N)$ representation):
\begin{align}
Z_{R}^\textrm{CS}=\langle W_{R} ( \CK)\rangle = \displaystyle\int \! [\mathcal{D}\mathcal{A}] \, e^{i S_{\rm CS}[\mathcal{A}]}\textrm{Tr}_R P  \exp \left(-\oint_{\CK} \mathcal{A}\right)\;. \label{path-integral representation of wilson loop VEVs}
\end{align}
We show in \cite[section 6]{GKMY_long} (by generalizing \cite{Lee:2013ida}) that in the supersymmetric localization
on $S^2\times M$ the 5d Wilson loop \eqref{5d_W} reduces to the 3d Wilson loop of \eqref{path-integral representation of wilson loop VEVs},
thus proving the equivalence of two Wilson lines.

We can evaluate the partition functions \eqref{Z_CS} and \eqref{path-integral representation of wilson loop VEVs} either by the state-integral model
(\cite{Dimofte:2011gm} for $N=2$, and \cite{Dimofte:2013iv} for $N>2$),
or the ``cluster partition function'' of \cite{Terashima:2013fg} (which is based on \cite{Terashima:2011qi,Terashima:2011xe,Nagao:2011aa}).
As discussed in \cite[sections 3 and 4]{GKMY_long},
in both cases the answer can be written as an overlap of two states inside a certain quantum mechanical system
with finite degrees of freedom:
\begin{align}
Z_{\rho}^{\rm cluster}=\langle \frakM_\a, C_I=0| \Diamond^{\otimes L}\rangle \;.
\label{Z_overlap}
\end{align}
Here $| \Diamond^{\otimes L}\rangle \in \mathcal{H}$ is a state representing $L$ `octahedra', or more physically $L$ free $\CN=2$ chiral multiplets.
The parameters $\frakM_\a$ are the same as in \eqref{co-dimension 2 defects as b.c. in CS theory-1},
and $C_I=0$ are the gluing constraints representing the gluing of the octahedra (or $\CN=2$ chiral multiplets).
The choice of $\rho$ is reflected in the octahedron structures as well as the choice of gluing equations $C_I=0$. The Hilbert space
$\mathcal{H}$ is obtained by quantizing the space of flat connections on $M$. In this framework, co-dimension 4 defects are obtained by inserting a line operator $\hat{W}_R(\CK)$ in between:
\begin{align}
Z_{R}^{\rm cluster}=\langle \frakM_\a, C_I=0| \, \hat{W}_R(\CK)\, | \Diamond^{\otimes L}\rangle \;.
\end{align}
Classically, the loop $W_R(\CK)$ can be computed by the snake rules of \cite{Dimofte:2013iv,FockGoncharovH}.
The general rule for the quantization of the loops is not known, however
we will provide a well-defined quantization rules of a class of loop operators in \cite{GKMY_long},
by extending the formalism of \cite{Terashima:2013fg} to incorporate Wilson lines.

{\bf 3d $\mathcal{N}=2$ Theory.}---Our defects can also be discussed in the context of the
3d $\CN=2$ theory $T_N[\hat{M}]$. In this theory, the two types of defects play different roles. A co-dimension 2 defect
fills the entire 3d, hence changes the 3d theory itself. We denote the resulting theory by $T_N[\hat{M}\backslash K, \rho]$.
By contrast a co-dimension 4 defect is a loop operator inside the 3d theory $T_N[\hat{M}]$ (or $T_N[\hat{M}\backslash K, \rho]$ if it co-exists with a co-dimension 2 defect).

Quantitatively, we can compute the $(S^3/\mathbb{Z}_k)_b$ \cite{Kapustin:2009kz,Gang:2009wy,Jafferis:2010un,Hama:2010av,Hama:2011ea,Imamura:2011wg,Benini:2011nc} or $(S^2 \times S^1)_q$ \cite{Kim:2009wb,Imamura:2011su} partition function of $T_N[\hat{M}\backslash K, \rho]$:
\begin{align}
Z_{\rho}^{\textrm{3d } \CN=2} = Z_{(S^3/\mathbb{Z}_k)_b \textrm{ or } (S^2 \times S^1)_q}\left[ T_N[\hat{M} \backslash K, \rho] \right]\;,
\end{align}
which is to be identified with the CS partition function \eqref{Z_CS}:
\begin{align}
Z_{\rho}^\textrm{CS}  = Z_{\rho}^\textrm{3d $\CN=2$} \;,
\label{main_eq}
\end{align}
 under the identification of the parameters (recall \eqref{Planck}):
\begin{align}
\begin{split}
&(S^3/\mathbb{Z}_k)_b: k \in \mathbb{Z}_{>0}, \quad\sigma =k \frac{1-b^2}{1+b^2} \in \mathbb{R} \textrm{ or } i \mathbb{R} \, . \\
&(S^2\times S^1)_{q=e^{\hbar}}:  k=0, \quad \sigma \in i \mathbb{R} \;.
\end{split}
  \label{3d-3d relation}
\end{align}
$\sigma$ being real or imaginary depends if $b\in \mathbb{R}$ \cite{Hama:2010av}  or $|b|=1$ \cite{Imamura:2011wg}. In fact, given the expression \eqref{Z_overlap}, we can reverse-engineer an Abelian gauge theory $T_N[\hatM\backslash K, \rho]$,
as has been done in \cite{Dimofte:2011gm,Dimofte:2013iv,Dimofte:2014zga} for maximal punctures and in our paper \cite{GKMY_long} for simple punctures.

It turns out that we run into problems for a co-dimension 4 defect, however.
As long as we use the Abelian gauge theory description, it is hard to understand why such a defect is labeled by
a representation $R$ \eqref{R}.
This problem is solved by the non-Abelian description of the $T_N[\hatM\backslash K, \rho]$ theory
given in \cite{Terashima:2011qi}, which works when $\rho$ is simple type.

In \cite{Terashima:2011qi}, the 3d theory is obtained as a duality domain wall between two 4d $\CN=2^{*}$ theories whose complexified gauge-coupling constants are related by an element of the S-duality group
$\varphi \in SL(2, \mathbb{Z})$.  Since the 4d $\CN=2^*$ theory is the 6d theory on a torus with a simple puncture,
we are restricted to the case where $\rho$ is simple.
The resulting 3d theory, which was denoted by $T_N[SU(N); \varphi]$ in \cite{Terashima:2011qi}, corresponds to a
3-manifold known as the mapping torus:
\begin{align}
&M= \left( \Sigma_{1,1} \times S^1 \right)_{\varphi}:  =  \{ (x,t)\in \Sigma_{1,1} \times [0,1] \}/\sim  \;,
\label{mapping_torus}
\end{align}
where  $\varphi$ is an element of $PSL(2, \mathbb{Z})$ and the equivalence relation $\sim$ is given by $(x,0)\sim (\varphi(x),1)$. In this non-abelian gauge theory description the co-dimension 4 defects are the Wilson lines
of the $\CN=2$ theory, so it is obvious why they carry a representation label \eqref{R}.
This resolves the paradox for the co-dimension 4 defects.
Unfortunately, for general cases
such a non-Abelian description of $T_N[\hatM\backslash K, \rho]$ theory is not known.

\begin{figure}
\centering\includegraphics[scale=0.13]{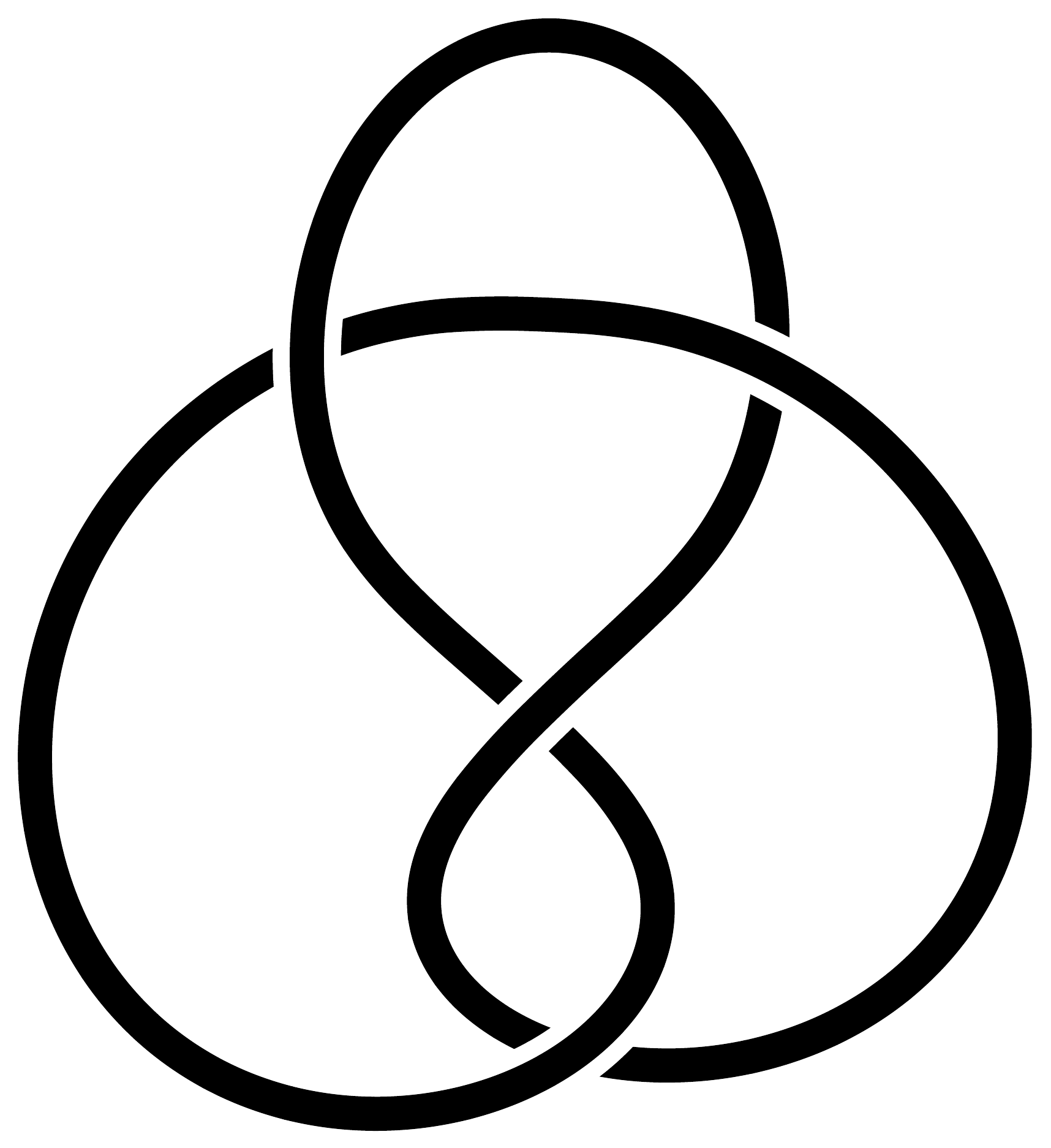}
\caption{The figure-eight knot inside $\hatM=S^3$.}
\label{fig.4_1}
\end{figure}

{\bf Quantitative Checks.}---Now that we have identified co-dimension 2 and 4
defects in various corners of \eqref{corners}, we can move to the quantitative checks of the
3d--3d correspondence, such as \eqref{main_eq} or its counterpart for co-dimension 4 defects.
Our companion paper \cite{GKMY_long} contains many such computations.
Let us here consider one example,
where we have a co-dimension 2 defect of simple type along the
figure-eight knot $\bf{4}_1$ in $\hatM=S^3$ (Fig.~\ref{fig.4_1}).

It is well-known that that this knot complement is hyperbolic, and can be triangulated by two
ideal tetrahedra. However, no state-integral model has been known in the literature for the case of a simple co-dimension 2 defect,
and we need an alternative approach \footnote{Interestingly, our result from the cluster partition function in \cite{GKMY_long} does give a new octahedron-like decomposition for the case with simple punctures. This suggests the possibility of extending the state-integral model construction to more general co-dimension 2 defects.}.
What helps us is that figure-eight knot complement is also a mapping torus \eqref{mapping_torus},
with $\varphi=\left(\begin{array}{cc} 2& 1\\ 1& 0 \end{array}\right) \in PSL(2, \mathbb{Z})$.

\begin{figure}
\begin{center}
   \includegraphics[width=.23\textwidth]{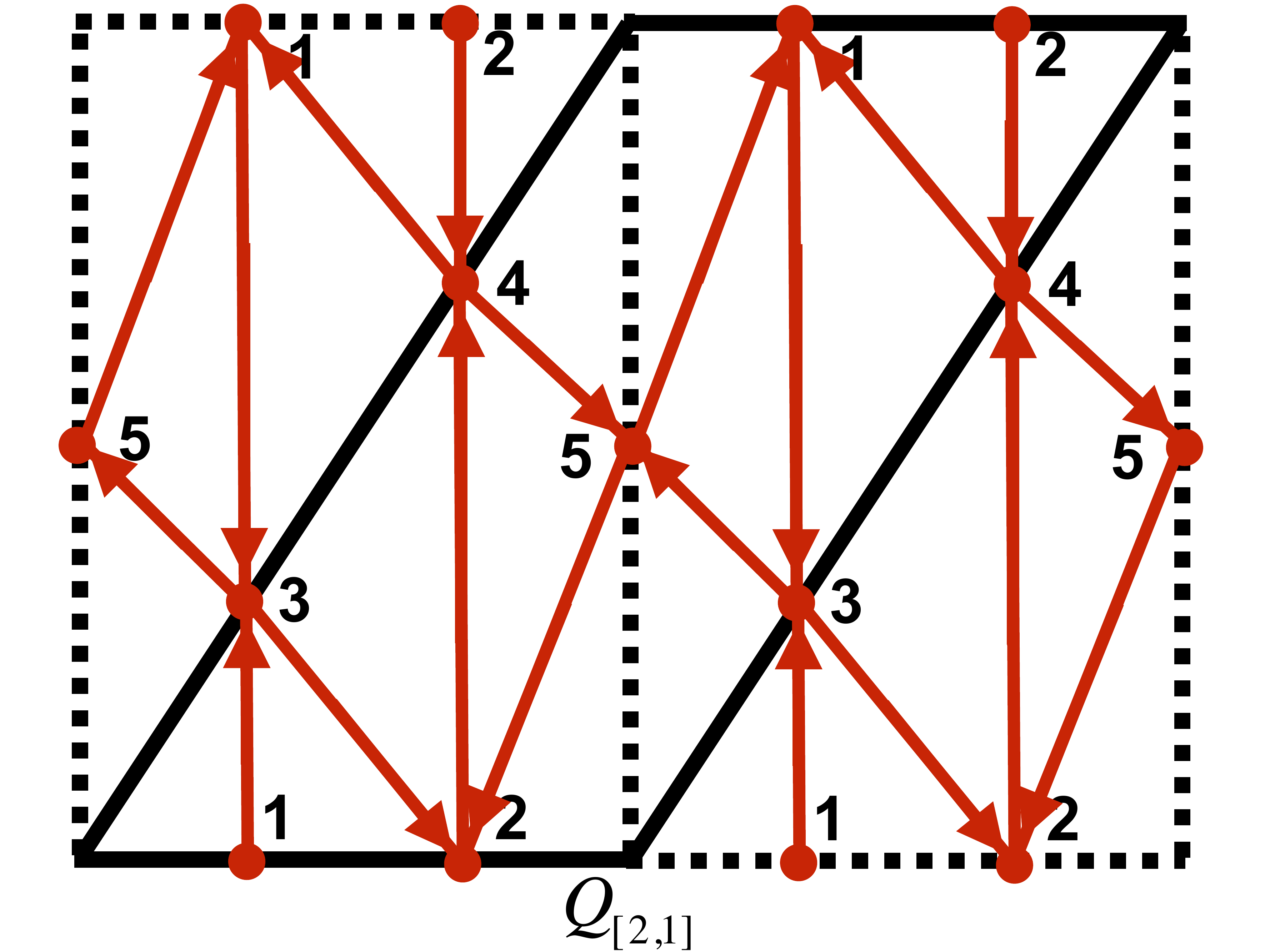}
   \end{center}
   \caption{The proposed quiver for a once-punctured torus, where we take $N=3$ and the puncture is of the simple type ($\rho=[2,1]$).
  The fundamental region of torus is chosen as the region surrounded by solid black  lines.}
    \label{fig:simple-once-punctured-quiver}
\end{figure}

We can now use two different methods. One is to use the cluster partition function of \cite{Terashima:2013fg,GKMY_long},
applied to the quiver of Fig.~\ref{fig:simple-once-punctured-quiver} for $N=3$ ({\it cf.} \cite{Xie:2012dw,FominP}). In \cite{GKMY_long} we have checked the consistency of the quiver, and found for example that the
mapping class group $PSL(2, \mathbb{Z})$ is indeed realized by a sequence of quiver mutations and permutations of the quiver vertices.

Another method is to use the $T[SU(N), \varphi]$ theory described previously.
As established in \cite{Terashima:2011qi},
the 3d $\mathcal{N}=2$ theory $T_{N=3}[S^{3}\setminus \mathbf{4_{1}},\rho=[2,1]]$ for the figure-eight knot complement can be then constructed from two copies of the $T[SU(3)]$ theory. The flavor symmetry group of $T[SU(3)]$ includes a factor $SU(3)_{\mathrm{bot}}\times SU(3)_{\mathrm{top}}$. First we define
\begin{align}
\begin{split}
&T\left[SU(3),\varphi=\mathbf{STS^{-1}T^{-1}}\right] \\
&\quad=T[SU(3),\mathbf{ST}]\odot T\left[SU(3),\mathbf{S^{-1}T^{-1}}\right] \;.
\end{split}
\end{align}
Here, the theories $T[SU(3),\mathbf{ST}]$ and $T[SU(3),\mathbf{S^{-1}T^{-1}}]$ are equivalent to $T[SU(3)]$ plus some additional CS terms and the operation
$\odot$ is defined by identifying $SU(3)_{\mathrm{top}}$ of the first factor with $SU(3)_{\mathrm{top}}$ of the second and then gauging. Finally, $T_{3}[S^{3}\backslash \mathbf{4_{1}},[2,1]]$ is constructed as
\begin{align}
T_{N=3}\left[S^{3}\backslash \mathbf{4_{1}},[2,1]\right]=\mathrm{Tr}\left(T\left[SU(3),\mathbf{STS^{-1}T^{-1}}\right]\right) \;,
\end{align}
where the operation $\mathrm{Tr}( \ )$ is defined by identifying the flavor symmetries $SU(3)_{\mathrm{bot}}$ and $SU(3)_{\mathrm{top}}$ of $T[SU(3),\varphi]$ and then gauging.

In both cases, we can straightforwardly compute their supersymmetric partition functions.
For the $(S^1\times S^2)_q$ partition function,  we have verified that the two methods mentioned above give
the same answer, at least up to certain orders in the $q$-expansion. For example,
{\footnotesize \begin{align}
\begin{split}
&Z_{(S^1\times S^2)_q}(m_\eta=0, \eta)  \\
&\qquad =1+ \left(2\eta+\frac{2}{\eta} \right) q^{\frac{3}{2}}+\left(8+2 \eta^2 + \frac{2}{\eta^2}\right) q^2
\\
& \qquad +\left(6 \eta+\frac{6}\eta \right)q^{\frac{5}{2}}+\left(2- 3 \eta^2- \frac{3}{\eta^2}\right)q^3+\ldots  \;,
\\
&Z_{(S^1\times S^2)_q}(m_\eta=1, \eta)  \\
& \qquad=   \left(\frac{1}{\eta^2}+\frac{1}{\eta} + \eta +\eta^2 \right) q+ \left(6 +3\eta +  \frac{3}{\eta}\right) q^2
\\
&\qquad + \left(-6 - \frac{1}{\eta^3}- \frac{3}{\eta^2}-\frac{5}{\eta}-5 \eta-3\eta^2 -\eta^3\right)q^3+\ldots \;,
\label{STSinvTinv}
\end{split}
\end{align}
}where $(\eta, m_{\eta})$ are the fugacity and the magnetic flux for a $U(1)$ flavor symmetry.
This symmetry corresponds geometrically to the puncture holonomy of the once-punctured torus,
or more physically to a part of the R-symmetry of the $\CN=4$ $T[SU(N); \varphi]$ theory, deforming the theory to  $\CN=2$.
The computation \eqref{STSinvTinv} is a highly non-trivial check for our proposed quiver (Fig.~\ref{fig:simple-once-punctured-quiver}),
as well as for the consistency check between Abelian and non-Abelian descriptions of $T_N[\hatM\backslash K, \rho]$ theory.

The paper \cite{GKMY_long} also contains quantitative results for the co-dimension 4 defects
from the $T[SU(N=2)]$ theory and the cluster partition function.

{\bf Large $N$.}---Finally, we can discuss the large $N$ limit. The $D=11$ supergravity background
takes the form of a warped product $AdS_4 \times \hat{M} \times \tilde{S}^4$ \cite{Gauntlett:2000ng,Donos:2010ax,Gang:2014qla,Gang:2014ema}:
\begin{align}
\begin{split}
&\dd s^2_{11} = l_{\rm P}^2 (2\pi N)^{\frac{2}{3}} (1+\sin^2\theta)^{\frac{1}{3}} \\
&\;\; \Bigg[ \dd s^2(AdS_4) + \dd s^2(H^3)
+ (1+\sin^2\theta)^{\frac{2}{3}} \dd s^2(\tilde{S}^4)
\Bigg] \;,
\label{metric}
\end{split}
\end{align}
where $l_{\rm P}$ is the $D=11$ Planck constant.
The warp factors depend on $\theta$, which is one of the coordinates of the squashed 4-sphere $\tilde{S}^4$.
In addition to that, $\tilde{S}^4$ contains a round $\tilde{S}^2$ which is fibered over $H^3$. Here
the $H^3$ is the local representation of the closed 3-manifold $\hatM=H^3/\Gamma$, where $\Gamma$ is a torsionless discrete subgroup of $PSL(2, \mathbb{C})$. $\hatM$ thus obtained has a finite volume, and it is free of orbifold singularities.

The supergravity computation gives, for the $S^3_b$ partition function with maximal defect,
\begin{align}
\log \left( Z_{\rho}^{\rm SUGRA} \right)= \frac{N^3}{12\pi } (b+b^{-1})^2\, \textrm{vol}(\hat{M}\backslash K)\;. \label{F_H3 for knot complement}
\end{align}
The co-dimension 2 maximal defect has the $N^3$ scaling, and back-reacts to the geometry, replacing the
geometry $\hatM$ by $\hatM\backslash K$.

A co-dimension 4 defect is described by the M2/M5-branes (recall \eqref{codim_4_brane}).
Our analysis in \cite{GKMY_long} shows the following:
the Wilson line in the fundamental representation ($R=\square$)
corresponds to a probe M2-brane filling the cycle $\CK$ inside $M$.
When we consider the Wilson in $K$-th anti-symmetric representation ($R=A_K$) with $K$ of order $N$, then M2-brane blows up
into the probe M5-brane ({\it cf.}\,\,\cite{Yamaguchi:2006tq,Gomis:2006sb,Assel:2012nf}).
In the latter case, the supergravity computation gives the leading large $N$ answer to be \cite{GKMY_long}
\be
\log \left( Z_{R}^{\rm SUGRA} \right) =  \frac{1+b^2}{2} N^2  \ell(\CK) \frac{K}{N}\left( 1-\frac{K}{N} \right) \;,
\label{codim4_action_general}
\ee
where $\ell(\CK)$ is the hyperbolic length of the knot $\CK$.
In \cite{GKMY_long} we reproduced the $b^{0}$ term of this result from the large $N$ analysis of the Chern-Simons theory.
We also computed the exact $S^3_{b=1}$ partition function of the $\mathrm{Tr}(T[SU(N), \varphi])$ theory ({\it cf.}\,\,\cite{Nishioka:2011dq,Benvenuti:2011ga,Gulotta:2011si,Assel:2012cp}).
More generally, these large $N$ results give fascinating predictions for the asymptotic behavior of
the perturbative CS partition functions and the cluster partition functions.
It would be interesting to explore the implications
of this large $N$ analysis in more depth.

{\bf Acknowledgements:} We would like to thank H. Chung, T. Dimofte, D.\ Xie and K.\ Yonekura for discussion.
The research of DG, MR and MY is supported by the WPI Initiative, MEXT, Japan.
DG is also supported by a Grant-in-Aid for Scientific Research on Innovative Areas 2303, MEXT.
MY is also supported by  JSPS Program for Advancing Strategic
International Networks to Accelerate the Circulation of Talented Researchers,
and by JSPS KAKENHI Grant 15K17634. MR acknowledges support from the Institute for
Advanced Study. NK acknowledges the sabbatical leave program of
Kyung Hee University.

\bibliographystyle{apsrev4-1}
\bibliography{w3c_letter}

\begin{thebibliography}{50}%
\makeatletter
\providecommand \@ifxundefined [1]{%
 \@ifx{#1\undefined}
}%
\providecommand \@ifnum [1]{%
 \ifnum #1\expandafter \@firstoftwo
 \else \expandafter \@secondoftwo
 \fi
}%
\providecommand \@ifx [1]{%
 \ifx #1\expandafter \@firstoftwo
 \else \expandafter \@secondoftwo
 \fi
}%
\providecommand \natexlab [1]{#1}%
\providecommand \enquote  [1]{``#1''}%
\providecommand \bibnamefont  [1]{#1}%
\providecommand \bibfnamefont [1]{#1}%
\providecommand \citenamefont [1]{#1}%
\providecommand \href@noop [0]{\@secondoftwo}%
\providecommand \href [0]{\begingroup \@sanitize@url \@href}%
\providecommand \@href[1]{\@@startlink{#1}\@@href}%
\providecommand \@@href[1]{\endgroup#1\@@endlink}%
\providecommand \@sanitize@url [0]{\catcode `\\12\catcode `\$12\catcode
  `\&12\catcode `\#12\catcode `\^12\catcode `\_12\catcode `\%12\relax}%
\providecommand \@@startlink[1]{}%
\providecommand \@@endlink[0]{}%
\providecommand \url  [0]{\begingroup\@sanitize@url \@url }%
\providecommand \@url [1]{\endgroup\@href {#1}{\urlprefix }}%
\providecommand \urlprefix  [0]{URL }%
\providecommand \Eprint [0]{\href }%
\providecommand \doibase [0]{http://dx.doi.org/}%
\providecommand \selectlanguage [0]{\@gobble}%
\providecommand \bibinfo  [0]{\@secondoftwo}%
\providecommand \bibfield  [0]{\@secondoftwo}%
\providecommand \translation [1]{[#1]}%
\providecommand \BibitemOpen [0]{}%
\providecommand \bibitemStop [0]{}%
\providecommand \bibitemNoStop [0]{.\EOS\space}%
\providecommand \EOS [0]{\spacefactor3000\relax}%
\providecommand \BibitemShut  [1]{\csname bibitem#1\endcsname}%
\let\auto@bib@innerbib\@empty
\bibitem [{\citenamefont {Gang}\ \emph {et~al.}()\citenamefont {Gang},
  \citenamefont {Kim}, \citenamefont {Romo},\ and\ \citenamefont
  {Yamazaki}}]{GKMY_long}%
  \BibitemOpen
  \bibfield  {author} {\bibinfo {author} {\bibfnamefont {D.}~\bibnamefont
  {Gang}}, \bibinfo {author} {\bibfnamefont {N.}~\bibnamefont {Kim}}, \bibinfo
  {author} {\bibfnamefont {M.}~\bibnamefont {Romo}}, \ and\ \bibinfo {author}
  {\bibfnamefont {M.}~\bibnamefont {Yamazaki}},\ }\href@noop {} {\enquote
  {\bibinfo {title} {{To Appear}},}\ }\BibitemShut {NoStop}%
\bibitem [{\citenamefont {Terashima}\ and\ \citenamefont
  {Yamazaki}(2011)}]{Terashima:2011qi}%
  \BibitemOpen
  \bibfield  {author} {\bibinfo {author} {\bibfnamefont {Y.}~\bibnamefont
  {Terashima}}\ and\ \bibinfo {author} {\bibfnamefont {M.}~\bibnamefont
  {Yamazaki}},\ }\href {\doibase 10.1007/JHEP08(2011)135} {\bibfield  {journal}
  {\bibinfo  {journal} {JHEP}\ }\textbf {\bibinfo {volume} {08}},\ \bibinfo
  {pages} {135} (\bibinfo {year} {2011})},\ \Eprint
  {http://arxiv.org/abs/1103.5748} {arXiv:1103.5748 [hep-th]} \BibitemShut
  {NoStop}%
\bibitem [{\citenamefont {Terashima}\ and\ \citenamefont
  {Yamazaki}(2013)}]{Terashima:2011xe}%
  \BibitemOpen
  \bibfield  {author} {\bibinfo {author} {\bibfnamefont {Y.}~\bibnamefont
  {Terashima}}\ and\ \bibinfo {author} {\bibfnamefont {M.}~\bibnamefont
  {Yamazaki}},\ }\href {\doibase 10.1103/PhysRevD.88.026011} {\bibfield
  {journal} {\bibinfo  {journal} {Phys. Rev.}\ }\textbf {\bibinfo {volume}
  {D88}},\ \bibinfo {pages} {026011} (\bibinfo {year} {2013})},\ \Eprint
  {http://arxiv.org/abs/1106.3066} {arXiv:1106.3066 [hep-th]} \BibitemShut
  {NoStop}%
\bibitem [{\citenamefont {Dimofte}\ \emph {et~al.}(2011)\citenamefont
  {Dimofte}, \citenamefont {Gukov},\ and\ \citenamefont
  {Hollands}}]{Dimofte:2010tz}%
  \BibitemOpen
  \bibfield  {author} {\bibinfo {author} {\bibfnamefont {T.}~\bibnamefont
  {Dimofte}}, \bibinfo {author} {\bibfnamefont {S.}~\bibnamefont {Gukov}}, \
  and\ \bibinfo {author} {\bibfnamefont {L.}~\bibnamefont {Hollands}},\ }\href
  {\doibase 10.1007/s11005-011-0531-8} {\bibfield  {journal} {\bibinfo
  {journal} {Lett. Math. Phys.}\ }\textbf {\bibinfo {volume} {98}},\ \bibinfo
  {pages} {225} (\bibinfo {year} {2011})},\ \Eprint
  {http://arxiv.org/abs/1006.0977} {arXiv:1006.0977 [hep-th]} \BibitemShut
  {NoStop}%
\bibitem [{\citenamefont {Dimofte}\ and\ \citenamefont
  {Gukov}(2013)}]{Dimofte:2011jd}%
  \BibitemOpen
  \bibfield  {author} {\bibinfo {author} {\bibfnamefont {T.}~\bibnamefont
  {Dimofte}}\ and\ \bibinfo {author} {\bibfnamefont {S.}~\bibnamefont
  {Gukov}},\ }\href {\doibase 10.1007/JHEP05(2013)109} {\bibfield  {journal}
  {\bibinfo  {journal} {JHEP}\ }\textbf {\bibinfo {volume} {05}},\ \bibinfo
  {pages} {109} (\bibinfo {year} {2013})},\ \Eprint
  {http://arxiv.org/abs/1106.4550} {arXiv:1106.4550 [hep-th]} \BibitemShut
  {NoStop}%
\bibitem [{\citenamefont {Dimofte}\ \emph {et~al.}(2014)\citenamefont
  {Dimofte}, \citenamefont {Gaiotto},\ and\ \citenamefont
  {Gukov}}]{Dimofte:2011ju}%
  \BibitemOpen
  \bibfield  {author} {\bibinfo {author} {\bibfnamefont {T.}~\bibnamefont
  {Dimofte}}, \bibinfo {author} {\bibfnamefont {D.}~\bibnamefont {Gaiotto}}, \
  and\ \bibinfo {author} {\bibfnamefont {S.}~\bibnamefont {Gukov}},\ }\href
  {\doibase 10.1007/s00220-013-1863-2} {\bibfield  {journal} {\bibinfo
  {journal} {Commun. Math. Phys.}\ }\textbf {\bibinfo {volume} {325}},\
  \bibinfo {pages} {367} (\bibinfo {year} {2014})},\ \Eprint
  {http://arxiv.org/abs/1108.4389} {arXiv:1108.4389 [hep-th]} \BibitemShut
  {NoStop}%
\bibitem [{\citenamefont {Cecotti}\ \emph {et~al.}(2011)\citenamefont
  {Cecotti}, \citenamefont {Cordova},\ and\ \citenamefont
  {Vafa}}]{Cecotti:2011iy}%
  \BibitemOpen
  \bibfield  {author} {\bibinfo {author} {\bibfnamefont {S.}~\bibnamefont
  {Cecotti}}, \bibinfo {author} {\bibfnamefont {C.}~\bibnamefont {Cordova}}, \
  and\ \bibinfo {author} {\bibfnamefont {C.}~\bibnamefont {Vafa}},\ }\href@noop
  {} {\  (\bibinfo {year} {2011})},\ \Eprint {http://arxiv.org/abs/1110.2115}
  {arXiv:1110.2115 [hep-th]} \BibitemShut {NoStop}%
\bibitem [{\citenamefont {Thurston}(8 79)}]{ThurstonLecture}%
  \BibitemOpen
  \bibfield  {author} {\bibinfo {author} {\bibfnamefont {W.~P.}\ \bibnamefont
  {Thurston}},\ }\href@noop {} {\enquote {\bibinfo {title} {The geometry and
  topology of three-manifolds},}\ } (\bibinfo {year} {1978-79})\BibitemShut
  {NoStop}%
\bibitem [{\citenamefont {Cordova}\ and\ \citenamefont
  {Jafferis}(2013)}]{Cordova:2013cea}%
  \BibitemOpen
  \bibfield  {author} {\bibinfo {author} {\bibfnamefont {C.}~\bibnamefont
  {Cordova}}\ and\ \bibinfo {author} {\bibfnamefont {D.~L.}\ \bibnamefont
  {Jafferis}},\ }\href@noop {} {\  (\bibinfo {year} {2013})},\ \Eprint
  {http://arxiv.org/abs/1305.2891} {arXiv:1305.2891 [hep-th]} \BibitemShut
  {NoStop}%
\bibitem [{\citenamefont {Lee}\ and\ \citenamefont
  {Yamazaki}(2013)}]{Lee:2013ida}%
  \BibitemOpen
  \bibfield  {author} {\bibinfo {author} {\bibfnamefont {S.}~\bibnamefont
  {Lee}}\ and\ \bibinfo {author} {\bibfnamefont {M.}~\bibnamefont {Yamazaki}},\
  }\href {\doibase 10.1007/JHEP12(2013)035} {\bibfield  {journal} {\bibinfo
  {journal} {JHEP}\ }\textbf {\bibinfo {volume} {12}},\ \bibinfo {pages} {035}
  (\bibinfo {year} {2013})},\ \Eprint {http://arxiv.org/abs/1305.2429}
  {arXiv:1305.2429 [hep-th]} \BibitemShut {NoStop}%
\bibitem [{\citenamefont {Yagi}(2013)}]{Yagi:2013fda}%
  \BibitemOpen
  \bibfield  {author} {\bibinfo {author} {\bibfnamefont {J.}~\bibnamefont
  {Yagi}},\ }\href {\doibase 10.1007/JHEP08(2013)017} {\bibfield  {journal}
  {\bibinfo  {journal} {JHEP}\ }\textbf {\bibinfo {volume} {1308}},\ \bibinfo
  {pages} {017} (\bibinfo {year} {2013})},\ \Eprint
  {http://arxiv.org/abs/1305.0291} {arXiv:1305.0291 [hep-th]} \BibitemShut
  {NoStop}%
\bibitem [{\citenamefont {Gaiotto}\ and\ \citenamefont
  {Witten}(2009{\natexlab{a}})}]{Gaiotto:2008sa}%
  \BibitemOpen
  \bibfield  {author} {\bibinfo {author} {\bibfnamefont {D.}~\bibnamefont
  {Gaiotto}}\ and\ \bibinfo {author} {\bibfnamefont {E.}~\bibnamefont
  {Witten}},\ }\href {\doibase 10.1007/s10955-009-9687-3} {\bibfield  {journal}
  {\bibinfo  {journal} {J. Statist. Phys.}\ }\textbf {\bibinfo {volume}
  {135}},\ \bibinfo {pages} {789} (\bibinfo {year} {2009}{\natexlab{a}})},\
  \Eprint {http://arxiv.org/abs/0804.2902} {arXiv:0804.2902 [hep-th]}
  \BibitemShut {NoStop}%
\bibitem [{\citenamefont {Gaiotto}\ and\ \citenamefont
  {Witten}(2009{\natexlab{b}})}]{Gaiotto:2008ak}%
  \BibitemOpen
  \bibfield  {author} {\bibinfo {author} {\bibfnamefont {D.}~\bibnamefont
  {Gaiotto}}\ and\ \bibinfo {author} {\bibfnamefont {E.}~\bibnamefont
  {Witten}},\ }\href {\doibase 10.4310/ATMP.2009.v13.n3.a5} {\bibfield
  {journal} {\bibinfo  {journal} {Adv. Theor. Math. Phys.}\ }\textbf {\bibinfo
  {volume} {13}},\ \bibinfo {pages} {721} (\bibinfo {year}
  {2009}{\natexlab{b}})},\ \Eprint {http://arxiv.org/abs/0807.3720}
  {arXiv:0807.3720 [hep-th]} \BibitemShut {NoStop}%
\bibitem [{\citenamefont {Bullimore}\ and\ \citenamefont
  {Kim}(2015)}]{Bullimore:2014upa}%
  \BibitemOpen
  \bibfield  {author} {\bibinfo {author} {\bibfnamefont {M.}~\bibnamefont
  {Bullimore}}\ and\ \bibinfo {author} {\bibfnamefont {H.-C.}\ \bibnamefont
  {Kim}},\ }\href {\doibase 10.1007/JHEP05(2015)048} {\bibfield  {journal}
  {\bibinfo  {journal} {JHEP}\ }\textbf {\bibinfo {volume} {05}},\ \bibinfo
  {pages} {048} (\bibinfo {year} {2015})},\ \Eprint
  {http://arxiv.org/abs/1412.3872} {arXiv:1412.3872 [hep-th]} \BibitemShut
  {NoStop}%
\bibitem [{\citenamefont {Yonekura}(2014)}]{Yonekura:2013mya}%
  \BibitemOpen
  \bibfield  {author} {\bibinfo {author} {\bibfnamefont {K.}~\bibnamefont
  {Yonekura}},\ }\href {\doibase 10.1007/JHEP01(2014)142} {\bibfield  {journal}
  {\bibinfo  {journal} {JHEP}\ }\textbf {\bibinfo {volume} {01}},\ \bibinfo
  {pages} {142} (\bibinfo {year} {2014})},\ \Eprint
  {http://arxiv.org/abs/1310.7943} {arXiv:1310.7943 [hep-th]} \BibitemShut
  {NoStop}%
\bibitem [{\citenamefont {Witten}(1991)}]{Witten:1989ip}%
  \BibitemOpen
  \bibfield  {author} {\bibinfo {author} {\bibfnamefont {E.}~\bibnamefont
  {Witten}},\ }\href {\doibase 10.1007/BF02099116} {\bibfield  {journal}
  {\bibinfo  {journal} {Commun. Math. Phys.}\ }\textbf {\bibinfo {volume}
  {137}},\ \bibinfo {pages} {29} (\bibinfo {year} {1991})}\BibitemShut
  {NoStop}%
\bibitem [{\citenamefont {Witten}(2011)}]{Witten:2010cx}%
  \BibitemOpen
  \bibfield  {author} {\bibinfo {author} {\bibfnamefont {E.}~\bibnamefont
  {Witten}},\ }\bibfield  {booktitle} {\emph {\bibinfo {booktitle}
  {{Chern-Simons gauge theory: 20 years after. Proceedings, Workshop, Bonn,
  Germany, August 3-7, 2009}}},\ }\href@noop {} {\bibfield  {journal} {\bibinfo
   {journal} {AMS/IP Stud. Adv. Math.}\ }\textbf {\bibinfo {volume} {50}},\
  \bibinfo {pages} {347} (\bibinfo {year} {2011})},\ \Eprint
  {http://arxiv.org/abs/1001.2933} {arXiv:1001.2933 [hep-th]} \BibitemShut
  {NoStop}%
\bibitem [{\citenamefont {Bergeron}\ \emph {et~al.}(2011)\citenamefont
  {Bergeron}, \citenamefont {Falbel},\ and\ \citenamefont {Guilloux}}]{BFG}%
  \BibitemOpen
  \bibfield  {author} {\bibinfo {author} {\bibfnamefont {N.}~\bibnamefont
  {Bergeron}}, \bibinfo {author} {\bibfnamefont {E.}~\bibnamefont {Falbel}}, \
  and\ \bibinfo {author} {\bibfnamefont {A.}~\bibnamefont {Guilloux}},\
  }\href@noop {} {\  (\bibinfo {year} {2011})},\ \Eprint
  {http://arxiv.org/abs/1101.2742} {arXiv:1101.2742 [math.GT]} \BibitemShut
  {NoStop}%
\bibitem [{\citenamefont {{Garoufalidis}}\ \emph {et~al.}(2011)\citenamefont
  {{Garoufalidis}}, \citenamefont {{Thurston}},\ and\ \citenamefont
  {{Zickert}}}]{2011arXiv1111.2828G}%
  \BibitemOpen
  \bibfield  {author} {\bibinfo {author} {\bibfnamefont {S.}~\bibnamefont
  {{Garoufalidis}}}, \bibinfo {author} {\bibfnamefont {D.~P.}\ \bibnamefont
  {{Thurston}}}, \ and\ \bibinfo {author} {\bibfnamefont {C.~K.}\ \bibnamefont
  {{Zickert}}},\ }\href@noop {} {\bibfield  {journal} {\bibinfo  {journal}
  {ArXiv e-prints}\ } (\bibinfo {year} {2011})},\ \Eprint
  {http://arxiv.org/abs/1111.2828} {arXiv:1111.2828 [math.GT]} \BibitemShut
  {NoStop}%
\bibitem [{\citenamefont {Garoufalidis}\ \emph {et~al.}(2015)\citenamefont
  {Garoufalidis}, \citenamefont {Goerner},\ and\ \citenamefont
  {Zickert}}]{GGZ}%
  \BibitemOpen
  \bibfield  {author} {\bibinfo {author} {\bibfnamefont {S.}~\bibnamefont
  {Garoufalidis}}, \bibinfo {author} {\bibfnamefont {M.}~\bibnamefont
  {Goerner}}, \ and\ \bibinfo {author} {\bibfnamefont {C.~K.}\ \bibnamefont
  {Zickert}},\ }\href {\doibase 10.2140/agt.2015.15.565} {\bibfield  {journal}
  {\bibinfo  {journal} {Algebr. Geom. Topol.}\ }\textbf {\bibinfo {volume}
  {15}},\ \bibinfo {pages} {565} (\bibinfo {year} {2015})}\BibitemShut
  {NoStop}%
\bibitem [{\citenamefont {Garoufalidis}\ and\ \citenamefont
  {Zickert}(2013)}]{Garoufalidis:2013upa}%
  \BibitemOpen
  \bibfield  {author} {\bibinfo {author} {\bibfnamefont {S.}~\bibnamefont
  {Garoufalidis}}\ and\ \bibinfo {author} {\bibfnamefont {C.~K.}\ \bibnamefont
  {Zickert}},\ }\href@noop {} {\  (\bibinfo {year} {2013})},\ \Eprint
  {http://arxiv.org/abs/1310.2497} {arXiv:1310.2497 [math.GT]} \BibitemShut
  {NoStop}%
\bibitem [{\citenamefont {Dimofte}\ \emph {et~al.}(2013)\citenamefont
  {Dimofte}, \citenamefont {Gabella},\ and\ \citenamefont
  {Goncharov}}]{Dimofte:2013iv}%
  \BibitemOpen
  \bibfield  {author} {\bibinfo {author} {\bibfnamefont {T.}~\bibnamefont
  {Dimofte}}, \bibinfo {author} {\bibfnamefont {M.}~\bibnamefont {Gabella}}, \
  and\ \bibinfo {author} {\bibfnamefont {A.~B.}\ \bibnamefont {Goncharov}},\
  }\href@noop {} {\  (\bibinfo {year} {2013})},\ \Eprint
  {http://arxiv.org/abs/1301.0192} {arXiv:1301.0192 [hep-th]} \BibitemShut
  {NoStop}%
\bibitem [{\citenamefont {Dimofte}(2013)}]{Dimofte:2011gm}%
  \BibitemOpen
  \bibfield  {author} {\bibinfo {author} {\bibfnamefont {T.}~\bibnamefont
  {Dimofte}},\ }\href {\doibase 10.4310/ATMP.2013.v17.n3.a1} {\bibfield
  {journal} {\bibinfo  {journal} {Adv. Theor. Math. Phys.}\ }\textbf {\bibinfo
  {volume} {17}},\ \bibinfo {pages} {479} (\bibinfo {year} {2013})},\ \Eprint
  {http://arxiv.org/abs/1102.4847} {arXiv:1102.4847 [hep-th]} \BibitemShut
  {NoStop}%
\bibitem [{\citenamefont {Terashima}\ and\ \citenamefont
  {Yamazaki}(2014)}]{Terashima:2013fg}%
  \BibitemOpen
  \bibfield  {author} {\bibinfo {author} {\bibfnamefont {Y.}~\bibnamefont
  {Terashima}}\ and\ \bibinfo {author} {\bibfnamefont {M.}~\bibnamefont
  {Yamazaki}},\ }\href {\doibase 10.1093/ptep/PTT115} {\bibfield  {journal}
  {\bibinfo  {journal} {PTEP}\ }\textbf {\bibinfo {volume} {023}},\ \bibinfo
  {pages} {B01} (\bibinfo {year} {2014})},\ \Eprint
  {http://arxiv.org/abs/1301.5902} {arXiv:1301.5902 [hep-th]} \BibitemShut
  {NoStop}%
\bibitem [{\citenamefont {Nagao}\ \emph {et~al.}(2011)\citenamefont {Nagao},
  \citenamefont {Terashima},\ and\ \citenamefont {Yamazaki}}]{Nagao:2011aa}%
  \BibitemOpen
  \bibfield  {author} {\bibinfo {author} {\bibfnamefont {K.}~\bibnamefont
  {Nagao}}, \bibinfo {author} {\bibfnamefont {Y.}~\bibnamefont {Terashima}}, \
  and\ \bibinfo {author} {\bibfnamefont {M.}~\bibnamefont {Yamazaki}},\
  }\href@noop {} {\  (\bibinfo {year} {2011})},\ \Eprint
  {http://arxiv.org/abs/1112.3106} {arXiv:1112.3106 [math.GT]} \BibitemShut
  {NoStop}%
\bibitem [{\citenamefont {{Fock}}\ and\ \citenamefont
  {{Goncharov}}(2003)}]{FockGoncharovH}%
  \BibitemOpen
  \bibfield  {author} {\bibinfo {author} {\bibfnamefont {V.~V.}\ \bibnamefont
  {{Fock}}}\ and\ \bibinfo {author} {\bibfnamefont {A.~B.}\ \bibnamefont
  {{Goncharov}}},\ }\href@noop {} {\  (\bibinfo {year} {2003})},\ \Eprint
  {http://arxiv.org/abs/math/0311149} {math/0311149} \BibitemShut {NoStop}%
\bibitem [{\citenamefont {Kapustin}\ \emph {et~al.}(2010)\citenamefont
  {Kapustin}, \citenamefont {Willett},\ and\ \citenamefont
  {Yaakov}}]{Kapustin:2009kz}%
  \BibitemOpen
  \bibfield  {author} {\bibinfo {author} {\bibfnamefont {A.}~\bibnamefont
  {Kapustin}}, \bibinfo {author} {\bibfnamefont {B.}~\bibnamefont {Willett}}, \
  and\ \bibinfo {author} {\bibfnamefont {I.}~\bibnamefont {Yaakov}},\ }\href
  {\doibase 10.1007/JHEP03(2010)089} {\bibfield  {journal} {\bibinfo  {journal}
  {JHEP}\ }\textbf {\bibinfo {volume} {03}},\ \bibinfo {pages} {089} (\bibinfo
  {year} {2010})},\ \Eprint {http://arxiv.org/abs/0909.4559} {arXiv:0909.4559
  [hep-th]} \BibitemShut {NoStop}%
\bibitem [{\citenamefont {Gang}(2009)}]{Gang:2009wy}%
  \BibitemOpen
  \bibfield  {author} {\bibinfo {author} {\bibfnamefont {D.}~\bibnamefont
  {Gang}},\ }\href@noop {} {\  (\bibinfo {year} {2009})},\ \Eprint
  {http://arxiv.org/abs/0912.4664} {arXiv:0912.4664 [hep-th]} \BibitemShut
  {NoStop}%
\bibitem [{\citenamefont {Jafferis}(2012)}]{Jafferis:2010un}%
  \BibitemOpen
  \bibfield  {author} {\bibinfo {author} {\bibfnamefont {D.~L.}\ \bibnamefont
  {Jafferis}},\ }\href {\doibase 10.1007/JHEP05(2012)159} {\bibfield  {journal}
  {\bibinfo  {journal} {JHEP}\ }\textbf {\bibinfo {volume} {05}},\ \bibinfo
  {pages} {159} (\bibinfo {year} {2012})},\ \Eprint
  {http://arxiv.org/abs/1012.3210} {arXiv:1012.3210 [hep-th]} \BibitemShut
  {NoStop}%
\bibitem [{\citenamefont {Hama}\ \emph
  {et~al.}(2011{\natexlab{a}})\citenamefont {Hama}, \citenamefont {Hosomichi},\
  and\ \citenamefont {Lee}}]{Hama:2010av}%
  \BibitemOpen
  \bibfield  {author} {\bibinfo {author} {\bibfnamefont {N.}~\bibnamefont
  {Hama}}, \bibinfo {author} {\bibfnamefont {K.}~\bibnamefont {Hosomichi}}, \
  and\ \bibinfo {author} {\bibfnamefont {S.}~\bibnamefont {Lee}},\ }\href
  {\doibase 10.1007/JHEP03(2011)127} {\bibfield  {journal} {\bibinfo  {journal}
  {JHEP}\ }\textbf {\bibinfo {volume} {03}},\ \bibinfo {pages} {127} (\bibinfo
  {year} {2011}{\natexlab{a}})},\ \Eprint {http://arxiv.org/abs/1012.3512}
  {arXiv:1012.3512 [hep-th]} \BibitemShut {NoStop}%
\bibitem [{\citenamefont {Hama}\ \emph
  {et~al.}(2011{\natexlab{b}})\citenamefont {Hama}, \citenamefont {Hosomichi},\
  and\ \citenamefont {Lee}}]{Hama:2011ea}%
  \BibitemOpen
  \bibfield  {author} {\bibinfo {author} {\bibfnamefont {N.}~\bibnamefont
  {Hama}}, \bibinfo {author} {\bibfnamefont {K.}~\bibnamefont {Hosomichi}}, \
  and\ \bibinfo {author} {\bibfnamefont {S.}~\bibnamefont {Lee}},\ }\href
  {\doibase 10.1007/JHEP05(2011)014} {\bibfield  {journal} {\bibinfo  {journal}
  {JHEP}\ }\textbf {\bibinfo {volume} {05}},\ \bibinfo {pages} {014} (\bibinfo
  {year} {2011}{\natexlab{b}})},\ \Eprint {http://arxiv.org/abs/1102.4716}
  {arXiv:1102.4716 [hep-th]} \BibitemShut {NoStop}%
\bibitem [{\citenamefont {Imamura}\ and\ \citenamefont
  {Yokoyama}(2012)}]{Imamura:2011wg}%
  \BibitemOpen
  \bibfield  {author} {\bibinfo {author} {\bibfnamefont {Y.}~\bibnamefont
  {Imamura}}\ and\ \bibinfo {author} {\bibfnamefont {D.}~\bibnamefont
  {Yokoyama}},\ }\href {\doibase 10.1103/PhysRevD.85.025015} {\bibfield
  {journal} {\bibinfo  {journal} {Phys. Rev.}\ }\textbf {\bibinfo {volume}
  {D85}},\ \bibinfo {pages} {025015} (\bibinfo {year} {2012})},\ \Eprint
  {http://arxiv.org/abs/1109.4734} {arXiv:1109.4734 [hep-th]} \BibitemShut
  {NoStop}%
\bibitem [{\citenamefont {Benini}\ \emph {et~al.}(2012)\citenamefont {Benini},
  \citenamefont {Nishioka},\ and\ \citenamefont {Yamazaki}}]{Benini:2011nc}%
  \BibitemOpen
  \bibfield  {author} {\bibinfo {author} {\bibfnamefont {F.}~\bibnamefont
  {Benini}}, \bibinfo {author} {\bibfnamefont {T.}~\bibnamefont {Nishioka}}, \
  and\ \bibinfo {author} {\bibfnamefont {M.}~\bibnamefont {Yamazaki}},\ }\href
  {\doibase 10.1103/PhysRevD.86.065015} {\bibfield  {journal} {\bibinfo
  {journal} {Phys. Rev.}\ }\textbf {\bibinfo {volume} {D86}},\ \bibinfo {pages}
  {065015} (\bibinfo {year} {2012})},\ \Eprint {http://arxiv.org/abs/1109.0283}
  {arXiv:1109.0283 [hep-th]} \BibitemShut {NoStop}%
\bibitem [{\citenamefont {Kim}(2009)}]{Kim:2009wb}%
  \BibitemOpen
  \bibfield  {author} {\bibinfo {author} {\bibfnamefont {S.}~\bibnamefont
  {Kim}},\ }\href {\doibase 10.1016/j.nuclphysb.2012.07.015,
  10.1016/j.nuclphysb.2009.06.025} {\bibfield  {journal} {\bibinfo  {journal}
  {Nucl. Phys.}\ }\textbf {\bibinfo {volume} {B821}},\ \bibinfo {pages} {241}
  (\bibinfo {year} {2009})},\ \bibinfo {note} {[Erratum: Nucl.
  Phys.B864,884(2012)]},\ \Eprint {http://arxiv.org/abs/0903.4172}
  {arXiv:0903.4172 [hep-th]} \BibitemShut {NoStop}%
\bibitem [{\citenamefont {Imamura}\ and\ \citenamefont
  {Yokoyama}(2011)}]{Imamura:2011su}%
  \BibitemOpen
  \bibfield  {author} {\bibinfo {author} {\bibfnamefont {Y.}~\bibnamefont
  {Imamura}}\ and\ \bibinfo {author} {\bibfnamefont {S.}~\bibnamefont
  {Yokoyama}},\ }\href {\doibase 10.1007/JHEP04(2011)007} {\bibfield  {journal}
  {\bibinfo  {journal} {JHEP}\ }\textbf {\bibinfo {volume} {04}},\ \bibinfo
  {pages} {007} (\bibinfo {year} {2011})},\ \Eprint
  {http://arxiv.org/abs/1101.0557} {arXiv:1101.0557 [hep-th]} \BibitemShut
  {NoStop}%
\bibitem [{\citenamefont {Dimofte}(2015)}]{Dimofte:2014zga}%
  \BibitemOpen
  \bibfield  {author} {\bibinfo {author} {\bibfnamefont {T.}~\bibnamefont
  {Dimofte}},\ }\href {\doibase 10.1007/s00220-015-2401-1} {\bibfield
  {journal} {\bibinfo  {journal} {Commun. Math. Phys.}\ }\textbf {\bibinfo
  {volume} {339}},\ \bibinfo {pages} {619} (\bibinfo {year} {2015})},\ \Eprint
  {http://arxiv.org/abs/1409.0857} {arXiv:1409.0857 [hep-th]} \BibitemShut
  {NoStop}%
\bibitem [{Note1()}]{Note1}%
  \BibitemOpen
  \bibinfo {note} {Interestingly, our result from the cluster partition
  function in \cite {GKMY_long} does give a new octahedron-like decomposition
  for the case with simple punctures. This suggests the possibility of
  extending the state-integral model construction to more general co-dimension
  2 defects.}\BibitemShut {Stop}%
\bibitem [{\citenamefont {Xie}(2012)}]{Xie:2012dw}%
  \BibitemOpen
  \bibfield  {author} {\bibinfo {author} {\bibfnamefont {D.}~\bibnamefont
  {Xie}},\ }\href@noop {} {\  (\bibinfo {year} {2012})},\ \Eprint
  {http://arxiv.org/abs/1203.4573} {arXiv:1203.4573 [hep-th]} \BibitemShut
  {NoStop}%
\bibitem [{\citenamefont {Fomin}\ and\ \citenamefont
  {Pylyavskyy}(2014)}]{FominP}%
  \BibitemOpen
  \bibfield  {author} {\bibinfo {author} {\bibfnamefont {S.}~\bibnamefont
  {Fomin}}\ and\ \bibinfo {author} {\bibfnamefont {P.}~\bibnamefont
  {Pylyavskyy}},\ }\href {\doibase 10.1073/pnas.1313068111} {\bibfield
  {journal} {\bibinfo  {journal} {Proc. Natl. Acad. Sci. USA}\ }\textbf
  {\bibinfo {volume} {111}},\ \bibinfo {pages} {9680} (\bibinfo {year}
  {2014})}\BibitemShut {NoStop}%
\bibitem [{\citenamefont {Gauntlett}\ \emph {et~al.}(2001)\citenamefont
  {Gauntlett}, \citenamefont {Kim},\ and\ \citenamefont
  {Waldram}}]{Gauntlett:2000ng}%
  \BibitemOpen
  \bibfield  {author} {\bibinfo {author} {\bibfnamefont {J.~P.}\ \bibnamefont
  {Gauntlett}}, \bibinfo {author} {\bibfnamefont {N.}~\bibnamefont {Kim}}, \
  and\ \bibinfo {author} {\bibfnamefont {D.}~\bibnamefont {Waldram}},\ }\href
  {\doibase 10.1103/PhysRevD.63.126001} {\bibfield  {journal} {\bibinfo
  {journal} {Phys. Rev.}\ }\textbf {\bibinfo {volume} {D63}},\ \bibinfo {pages}
  {126001} (\bibinfo {year} {2001})},\ \Eprint
  {http://arxiv.org/abs/hep-th/0012195} {arXiv:hep-th/0012195 [hep-th]}
  \BibitemShut {NoStop}%
\bibitem [{\citenamefont {Donos}\ \emph {et~al.}(2010)\citenamefont {Donos},
  \citenamefont {Gauntlett}, \citenamefont {Kim},\ and\ \citenamefont
  {Varela}}]{Donos:2010ax}%
  \BibitemOpen
  \bibfield  {author} {\bibinfo {author} {\bibfnamefont {A.}~\bibnamefont
  {Donos}}, \bibinfo {author} {\bibfnamefont {J.~P.}\ \bibnamefont
  {Gauntlett}}, \bibinfo {author} {\bibfnamefont {N.}~\bibnamefont {Kim}}, \
  and\ \bibinfo {author} {\bibfnamefont {O.}~\bibnamefont {Varela}},\ }\href
  {\doibase 10.1007/JHEP12(2010)003} {\bibfield  {journal} {\bibinfo  {journal}
  {JHEP}\ }\textbf {\bibinfo {volume} {12}},\ \bibinfo {pages} {003} (\bibinfo
  {year} {2010})},\ \Eprint {http://arxiv.org/abs/1009.3805} {arXiv:1009.3805
  [hep-th]} \BibitemShut {NoStop}%
\bibitem [{\citenamefont {Gang}\ \emph {et~al.}(2014)\citenamefont {Gang},
  \citenamefont {Kim},\ and\ \citenamefont {Lee}}]{Gang:2014qla}%
  \BibitemOpen
  \bibfield  {author} {\bibinfo {author} {\bibfnamefont {D.}~\bibnamefont
  {Gang}}, \bibinfo {author} {\bibfnamefont {N.}~\bibnamefont {Kim}}, \ and\
  \bibinfo {author} {\bibfnamefont {S.}~\bibnamefont {Lee}},\ }\href {\doibase
  10.1016/j.physletb.2014.04.051} {\bibfield  {journal} {\bibinfo  {journal}
  {Phys. Lett.}\ }\textbf {\bibinfo {volume} {B733}},\ \bibinfo {pages} {316}
  (\bibinfo {year} {2014})},\ \Eprint {http://arxiv.org/abs/1401.3595}
  {arXiv:1401.3595 [hep-th]} \BibitemShut {NoStop}%
\bibitem [{\citenamefont {Gang}\ \emph {et~al.}(2015)\citenamefont {Gang},
  \citenamefont {Kim},\ and\ \citenamefont {Lee}}]{Gang:2014ema}%
  \BibitemOpen
  \bibfield  {author} {\bibinfo {author} {\bibfnamefont {D.}~\bibnamefont
  {Gang}}, \bibinfo {author} {\bibfnamefont {N.}~\bibnamefont {Kim}}, \ and\
  \bibinfo {author} {\bibfnamefont {S.}~\bibnamefont {Lee}},\ }\href {\doibase
  10.1007/JHEP04(2015)091} {\bibfield  {journal} {\bibinfo  {journal} {JHEP}\
  }\textbf {\bibinfo {volume} {04}},\ \bibinfo {pages} {091} (\bibinfo {year}
  {2015})},\ \Eprint {http://arxiv.org/abs/1409.6206} {arXiv:1409.6206
  [hep-th]} \BibitemShut {NoStop}%
\bibitem [{\citenamefont {Yamaguchi}(2006)}]{Yamaguchi:2006tq}%
  \BibitemOpen
  \bibfield  {author} {\bibinfo {author} {\bibfnamefont {S.}~\bibnamefont
  {Yamaguchi}},\ }\href {\doibase 10.1088/1126-6708/2006/05/037} {\bibfield
  {journal} {\bibinfo  {journal} {JHEP}\ }\textbf {\bibinfo {volume} {05}},\
  \bibinfo {pages} {037} (\bibinfo {year} {2006})},\ \Eprint
  {http://arxiv.org/abs/hep-th/0603208} {arXiv:hep-th/0603208 [hep-th]}
  \BibitemShut {NoStop}%
\bibitem [{\citenamefont {Gomis}\ and\ \citenamefont
  {Passerini}(2006)}]{Gomis:2006sb}%
  \BibitemOpen
  \bibfield  {author} {\bibinfo {author} {\bibfnamefont {J.}~\bibnamefont
  {Gomis}}\ and\ \bibinfo {author} {\bibfnamefont {F.}~\bibnamefont
  {Passerini}},\ }\href {\doibase 10.1088/1126-6708/2006/08/074} {\bibfield
  {journal} {\bibinfo  {journal} {JHEP}\ }\textbf {\bibinfo {volume} {08}},\
  \bibinfo {pages} {074} (\bibinfo {year} {2006})},\ \Eprint
  {http://arxiv.org/abs/hep-th/0604007} {arXiv:hep-th/0604007 [hep-th]}
  \BibitemShut {NoStop}%
\bibitem [{\citenamefont {Assel}\ \emph {et~al.}(2014)\citenamefont {Assel},
  \citenamefont {Estes},\ and\ \citenamefont {Yamazaki}}]{Assel:2012nf}%
  \BibitemOpen
  \bibfield  {author} {\bibinfo {author} {\bibfnamefont {B.}~\bibnamefont
  {Assel}}, \bibinfo {author} {\bibfnamefont {J.}~\bibnamefont {Estes}}, \ and\
  \bibinfo {author} {\bibfnamefont {M.}~\bibnamefont {Yamazaki}},\ }\href
  {\doibase 10.1007/s00023-013-0249-5} {\bibfield  {journal} {\bibinfo
  {journal} {Annales Henri Poincare}\ }\textbf {\bibinfo {volume} {15}},\
  \bibinfo {pages} {589} (\bibinfo {year} {2014})},\ \Eprint
  {http://arxiv.org/abs/1212.1202} {arXiv:1212.1202 [hep-th]} \BibitemShut
  {NoStop}%
\bibitem [{\citenamefont {Nishioka}\ \emph {et~al.}(2011)\citenamefont
  {Nishioka}, \citenamefont {Tachikawa},\ and\ \citenamefont
  {Yamazaki}}]{Nishioka:2011dq}%
  \BibitemOpen
  \bibfield  {author} {\bibinfo {author} {\bibfnamefont {T.}~\bibnamefont
  {Nishioka}}, \bibinfo {author} {\bibfnamefont {Y.}~\bibnamefont {Tachikawa}},
  \ and\ \bibinfo {author} {\bibfnamefont {M.}~\bibnamefont {Yamazaki}},\
  }\href {\doibase 10.1007/JHEP08(2011)003} {\bibfield  {journal} {\bibinfo
  {journal} {JHEP}\ }\textbf {\bibinfo {volume} {08}},\ \bibinfo {pages} {003}
  (\bibinfo {year} {2011})},\ \Eprint {http://arxiv.org/abs/1105.4390}
  {arXiv:1105.4390 [hep-th]} \BibitemShut {NoStop}%
\bibitem [{\citenamefont {Benvenuti}\ and\ \citenamefont
  {Pasquetti}(2012)}]{Benvenuti:2011ga}%
  \BibitemOpen
  \bibfield  {author} {\bibinfo {author} {\bibfnamefont {S.}~\bibnamefont
  {Benvenuti}}\ and\ \bibinfo {author} {\bibfnamefont {S.}~\bibnamefont
  {Pasquetti}},\ }\href {\doibase 10.1007/JHEP05(2012)099} {\bibfield
  {journal} {\bibinfo  {journal} {JHEP}\ }\textbf {\bibinfo {volume} {05}},\
  \bibinfo {pages} {099} (\bibinfo {year} {2012})},\ \Eprint
  {http://arxiv.org/abs/1105.2551} {arXiv:1105.2551 [hep-th]} \BibitemShut
  {NoStop}%
\bibitem [{\citenamefont {Gulotta}\ \emph {et~al.}(2011)\citenamefont
  {Gulotta}, \citenamefont {Herzog},\ and\ \citenamefont
  {Pufu}}]{Gulotta:2011si}%
  \BibitemOpen
  \bibfield  {author} {\bibinfo {author} {\bibfnamefont {D.~R.}\ \bibnamefont
  {Gulotta}}, \bibinfo {author} {\bibfnamefont {C.~P.}\ \bibnamefont {Herzog}},
  \ and\ \bibinfo {author} {\bibfnamefont {S.~S.}\ \bibnamefont {Pufu}},\
  }\href {\doibase 10.1007/JHEP12(2011)077} {\bibfield  {journal} {\bibinfo
  {journal} {JHEP}\ }\textbf {\bibinfo {volume} {12}},\ \bibinfo {pages} {077}
  (\bibinfo {year} {2011})},\ \Eprint {http://arxiv.org/abs/1105.2817}
  {arXiv:1105.2817 [hep-th]} \BibitemShut {NoStop}%
\bibitem [{\citenamefont {Assel}\ \emph {et~al.}(2012)\citenamefont {Assel},
  \citenamefont {Estes},\ and\ \citenamefont {Yamazaki}}]{Assel:2012cp}%
  \BibitemOpen
  \bibfield  {author} {\bibinfo {author} {\bibfnamefont {B.}~\bibnamefont
  {Assel}}, \bibinfo {author} {\bibfnamefont {J.}~\bibnamefont {Estes}}, \ and\
  \bibinfo {author} {\bibfnamefont {M.}~\bibnamefont {Yamazaki}},\ }\href
  {\doibase 10.1007/JHEP09(2012)074} {\bibfield  {journal} {\bibinfo  {journal}
  {JHEP}\ }\textbf {\bibinfo {volume} {09}},\ \bibinfo {pages} {074} (\bibinfo
  {year} {2012})},\ \Eprint {http://arxiv.org/abs/1206.2920} {arXiv:1206.2920
  [hep-th]} \BibitemShut {NoStop}%
\end{thebibliography}%

\end{document}